\documentclass[twocolumn,english,superscriptaddress,prb]{revtex4-1}
\usepackage{color}
\usepackage{bm}
\usepackage{amstext}
\usepackage{amssymb}
\usepackage{graphicx}
\usepackage{commath}
\usepackage{dsfont}
\usepackage{mathtools}
\usepackage{amsmath,mathrsfs}

\DeclareMathOperator{\sgn}{sgn}

\renewcommand{\vec}[1]{\mathbf{\boldsymbol{#1}}}

\usepackage{physics}

\begin{document}

\title{Reconstructing quantum molecular rotor ground states}

\author{Isaac J.S. De Vlugt}
\affiliation{Department of Physics and Astronomy, University of Waterloo, Waterloo, ON N2L 3G1}
\affiliation{Perimeter Institute for Theoretical Physics, Waterloo, ON N2L 2Y5}

\author{Dmitri Iouchtchenko}
\affiliation{Department of Chemistry, University of Waterloo, Waterloo, ON N2L 3G1}

\author{Ejaaz Merali}
\affiliation{Department of Physics and Astronomy, University of Waterloo, Waterloo, ON N2L 3G1}
\affiliation{Perimeter Institute for Theoretical Physics, Waterloo, ON N2L 2Y5}

\author{Pierre-Nicholas Roy}
\affiliation{Department of Chemistry, University of Waterloo, Waterloo, ON N2L 3G1}

\author{Roger G. Melko}
\affiliation{Department of Physics and Astronomy, University of Waterloo, Waterloo, ON N2L 3G1}
\affiliation{Perimeter Institute for Theoretical Physics, Waterloo, ON N2L 2Y5}

\date{\today}

\begin{abstract}

    Nanomolecular assemblies of C$_{60}$ can be synthesized to enclose dipolar molecules.  The low-temperature states of such endofullerenes
    are described by quantum mechanical rotors, which are candidates for quantum information devices with higher-dimensional local Hilbert spaces.
    The experimental exploration of endofullerene arrays comes at a time when machine learning techniques are rapidly being adopted to characterize,
    verify, and reconstruct quantum states from measurement data.  In this paper, we develop a strategy for reconstructing the ground state of
    chains of dipolar rotors using restricted Boltzmann machines (RBMs) adapted to train on data from higher-dimensional Hilbert spaces.  We demonstrate accurate
    generation of energy expectation values from an RBM trained on data in the free-rotor eigenstate basis, and explore the learning resources required for various
    chain lengths and dipolar interaction strengths.  Finally, we show evidence for fundamental limitations in the accuracy achievable by RBMs due to the
    difficulty in imposing symmetries in the sampling procedure.  We discuss possible avenues to overcome this limitation in the future, including
    the further development of autoregressive models such as recurrent neural networks for the purposes of quantum state reconstruction.

\end{abstract}

\maketitle

\section{Introduction} \label{sec:intro}

The upcoming decade will usher in an era of increasing hybridization between two rapidly advancing technologies: artificial intelligence and near-term quantum devices.  The abundance of data produced by today's quantum hardware has driven a number of machine learning innovations for reconstructing the underlying
quantum state of a system from data sets comprised of projective measurements on individual qubits.  A leading strategy uses generative models to learn and
represent the quantum state
\cite{torlaiLearningThermodynamicsBoltzmann2016,Microsoft_Tomo,torlaiNeuralNetworkQuantumState2018, Kappen,Torlai_ARCMP,beachQuCumberWavefunctionReconstruction2019}.
This approach has culminated in the recent demonstration of the reconstruction of a Rydberg atom quantum simulator from experimental data\cite{RydergReconstruction}.
Methods for quantum state reconstruction with generative models are advancing rapidly, and are capable of learning both pure state wavefunctions
and mixed state density matrices\cite{torlaiLatentSpacePurification2018,beachQuCumberWavefunctionReconstruction2019,NNtomoNori}.  Reconstruction is driven by projective qubit measurement data in a variety of bases, or more general positive operator valued measures (POVMs)
\cite{carrasquillaReconstructingQuantumStates2018}.

Although the potential of quantum hardware comprised of qubits has yet to be fully harnessed, the search for other candidates as
a foundation for quantum computers also continues in earnest.
Generalizations from two-dimensional local Hilbert spaces (qubits) to $d$-dimensional ones (qudits), or to continuous variables, can yield a significantly different platform
on which to devise quantum information protocols.  Particularly intriguing are recent proposals to harness the infinite-dimensional Hilbert space of rotational
states of a rigid body\cite{GKP2001}.  Noise-robust embeddings of finite-dimensional systems within this infinite-dimensional Hilbert space are possible, including
error-correcting ``molecular'' codes\cite{albertRobustEncodingQubit2019}.

A promising experimental platform to provide access to rotational quantum degrees of freedom is that of nanomolecular assemblies (NMAs) of endofullerenes.
Endofullerenes consist of a molecule encapsulated inside a carbon cage such as the buckminsterfullerene C$_{60}$ and can be synthesized using molecular surgery techniques \cite{komatsu2005encapsulation}.
It was recently suggested that NMAs composed of linear arrays of molecular endofullerenes could be used as quantum information devices \cite{halversonQuantifyingEntanglementRotor2018}.
These linear arrays can be synthesized using nanotechnology tools such as nanotube peapods \cite{berber2002microscopic}.
An additional handle on the properties of the NMAs is to use their electronic degrees of freedom in order to enhance transport \cite{xuElectricalDrivenTransportEndohedral2013}, or enable field-effect transistor behaviour \cite{leeNanoMemoryDevicesSingle2003}.

A more tunable medium is provided by adding the presence of dipolar molecules
within the C$_{60}$ cages, which gives discrete rotational quantum numbers appropriate for quantum information processing. One advantage of using NMAs of endofullerenes for quantum information processing is in their ability to retain quantum effects at relatively high temperatures, compared to some current qubit-based devices \cite{krachmalnicoffDipolarEndofullereneHF2016}.
The presence of dipole-dipole interactions between neighbouring molecules leads to coupling between the rotors and strong correlations in the system \cite{xuGeneralSelectionRule2015,felkerElectricDipoleCoupledTextsubscript2OTextsubscript602017}.
These correlations lead to entanglement between the rotors.

In order to study these properties, one must first calculate the many-body ground states of the NMA.
This can be achieved using exact diagonalization (ED) techniques along with basis truncation \cite{halversonQuantifyingEntanglementRotor2018},
an approach that however scales exponentially with system size.
A scalable alternative is the density matrix renormalization group (DMRG)
\cite{whiteDensityMatrixFormulation1992, whiteDensityMatrixAlgorithms1993, schollwoeckDensityMatrixRenormalizationGroup2011}, which has recently been extended
to calculate the ground state of large chains of coupled dipolar rotors\cite{iouchtchenkoGroundStatesLinear2018}.
Although these are first steps towards investigating the practicality of NMAs of endofullerenes as quantum hardware, an exploration of techniques for efficient state reconstruction using modern numerical methods is an obvious additional step in their development.

In this paper, we construct a microscopic simulation of an NMA of endofullerenes.  Using a model Hamiltonian consisting of rigid rotors with dipole-dipole interactions, the ground state wavefunction is studied using ED and DMRG.
From these simulations we observe a non-trivial sign structure of the ground state wavefunction in the free-rotor eigenstate basis of the rotating dipolar molecule.
Despite this, we are able to reconstruct states using standard machine learning generative models solely in this computational basis.
We demonstrate the procedure using a restricted Boltzmann machine (RBM), a
stochastic neural network comprised of two layers connected by weight parameters.
While an RBM commonly has binomial units in its visible layer, in our case it is generalized to have a
visible layer consisting of multinomial units, representing the angular momentum Hilbert space of each rotor.
Synthetic data is obtained through an exact sampling procedure on the DMRG wavefunction, and is used to
train the RBM parameters through a stochastic gradient descent algorithm.
Through this procedure, we are able to demonstrate accurate
reconstruction of the ground state energy for systems of up to 8 endofullerene sites.
We further characterize the reconstruction procedure by investigating the hidden layer size and input dataset size required to obtain an energy difference equal to 5\% of the first excitation gap.

\section{Nanomolecular Assemblies of Endofullerenes} \label{sec:rotor_theory}

\subsection{Hamiltonian and Hilbert Space} \label{sec:Ham_Hilb}

The Hamiltonian describing a linear chain of $N$ dipolar molecules that are equally spaced a distance $r$ apart includes, to a first approximation, the rotational energy of each molecule with a rotational constant $B$, and Coulombic dipole-dipole interactions with a dipole moment $\mu$:
\begin{equation} \label{eq:hamiltonian_full}
    \hat{H} = \frac{B}{\hbar^2} \sum_{i=1}^N\hat{\ell}_{i}^{2} + \frac{\mu^2}{4 \pi \epsilon_0 r^3} \sum_{i<j}\hat{V}_{ij},
\end{equation}
where
\begin{equation}
    \hat{\ell}_{i}^{2} \ket{\ell_j m_j} = \hbar^2 \ell_i(\ell_i + 1) \delta_{ij} \ket{\ell_i m_i},
\end{equation}
and
\begin{equation}
    \hat{V}_{ij} = \frac{\hat{x}_{i}\hat{x}_{j} + \hat{y}_{i}\hat{y}_{j} - 2\hat{z}_{i}\hat{z}_{j}}{\vert i-j \vert^3}.
\end{equation}
Without loss of generality, we can write this Hamiltonian in dimensionless form as\cite{iouchtchenkoGroundStatesLinear2018}
\begin{equation} \label{eq:hamiltonian}
    \frac{\hat{H}}{B} = \hat{K} + \frac{1}{R^3} \hat{V} = \sum_{i=1}^N\frac{\hat{\ell}_i^2}{\hbar^2} + \frac{1}{R^3} \sum_{i<j}\hat{V}_{ij},
\end{equation}
where,
\begin{equation}
    R = r\left(\frac{4 \pi \epsilon_0 B}{\mu^2}\right)^{1/3},
\end{equation}
is a dimensionless quantity. When each dipolar molecule is encapsulated by a fullerene (forming an endofullerene), the resulting screening of interactions can be modeled simply by reducing $\mu$. For example, the dipole moment of HF is reduced from
1.8265 Debye in the gas phase, to an effective 0.45 Debye when encapsulated in C$_{60}$\cite{krachmalnicoffDipolarEndofullereneHF2016}.

For dipolar diatomic molecules such as HF trapped in C$_{60}$, it has been shown that computed low lying eigenstates using \textit{ab initio} methods exhibit small coupling between molecular rotations and translations \cite{kaluginaPotentialEnergy2017}. These computed eigenstates are in agreement with experimentally measured spectra \cite{krachmalnicoffDipolarEndofullereneHF2016}.
Moreover, given that diatomic vibrations are at a much higher energy, molecules can be approximated as being in their ground vibrational state and behaving like rigid rotors. With regards to interactions between the dipolar molecule and the C$_{60}$ cage, it has also been shown that HF behaves as a nearly free rotor in C$_{60}$ (with a renormalized rotational constant of $18.523 \, \mathrm{cm}^{-1}$) due to the icosahedral symmetry of the cage \cite{kaluginaPotentialEnergy2017}.

For realistic assemblies of endofullerenes, the rotational energy is much larger than the strength of the Coulombic dipole-dipole interaction ($R \gg 1$).
For example, if the adjacent rotor distance is taken to be the diameter of C$_{60}$, 19.0 bohr\cite{felkerElectricDipoleCoupledTextsubscript2OTextsubscript602017}, and given the renormalized rotational constant and screened dipole moment of HF inside C$_{60}$, we find $R \approx 2.64$.
It is therefore a natural choice for the single-rotor reference basis to be the free-rotor eigenstate basis, $\ket{\ell_i m_i}$.
For convenience, we combine the angular momentum of a single rotor, $\ell_i$, and its projection, $m_i$, into one zero-indexed integer label, $\sigma_i$:
\begin{equation} \label{eq:ref_basis}
    \begin{aligned}
        \bigotimes_{i=1}^{N}\ket{\ell_i m_i} & = \ket{\ell_1 m_1, \ell_2 m_2, ... , \ell_N m_N} \\
                                             & = \ket{\sigma_1, \sigma_2, ... , \sigma_N}
        = \ket{\bm{\sigma}}.
    \end{aligned}
\end{equation}
Note that since the Hamiltonian in Eq.~\eqref{eq:hamiltonian} commutes with the $z$ projection of the total angular momentum, $\sum_{i=1}^N \hat{\ell}_{i,z}$, and with $\bigotimes_{i=1}^{N} \hat{\pi}_i$ (where $\hat{\pi}_i$ is the parity operator that inverts the $z_i$ coordinate in the position representation: $\hat{\pi}_i \ket{x_i \, y_i \, z_i} = \ket{x_i \, y_i \, {-z_i}}$), it conserves the total $m$ value $m = \sum_{i=1}^N m_i$ and the total $\ell$ parity $\ell_\mathrm{p} \equiv \sum_{i=1}^N \ell_i \pmod{2}$, respectively.

Given that the local Hilbert space of each rotor is infinite-dimensional, a truncation of the Hilbert space is required in order to perform numerical studies.
The truncation scheme adopted in this article is to cap the available angular momentum of each rotor to a maximum value, $\ell_{\mathrm{max}}$.
Thus, the local Hilbert space size, $\mathcal{D}$, becomes,
\begin{equation} \label{eq:local_hilbert}
    \mathcal{D} = \sum_{\ell = 0}^{\ell_{\mathrm{max}}} \left(2\ell + 1\right) = \left(\ell_{\mathrm{max}} + 1\right)^2 ,
\end{equation}
and the full Hilbert space size of the system of $N$ endofullerenes is $\mathcal{D}^N$.
As $R$ decreases (Coulombic interactions increase), $\ell_{\mathrm{max}}$ must be increased in order to faithfully represent the contribution of the dipole-dipole interaction and, therefore, the ground state of the system.

\subsection{Ground State Wavefunction Sign Structure} \label{sec:sign_structure}

We explore the use of RBMs in reconstructing the ground state wavefunction of Eq.~\eqref{eq:hamiltonian}
under the assumption that we have access to data obtained in the free-rotor eigenstate basis.
In this data-driven setting, the Hamiltonian itself is not used to determine the ground state, e.g.~through exact diagonalization,
variational Monte Carlo, or some other method.
Rather, it is assumed that an experimental system such as an endofullerene NMA is available, from which projective measurement data is produced
which enables a reconstruction.
Nonetheless, any {\it a priori} knowledge of the experimental Hamiltonian can be taken advantage of to increase the efficiency of the reconstruction.
In particular, when applying the RBM (or any generative model) for this purpose, it is crucial to be aware of the sign structure of the
wavefunction in the computational basis, since this has a direct impact on the complexity of the machine learning architecture required.
If the ground state wavefunction is real and positive, it can trivially be equated to the square root of the probability density,
and this simplifies the generative modeling drastically \cite{beachQuCumberWavefunctionReconstruction2019}.

The well-known Perron-Frobenius theorem gives a sufficient (but not necessary) criterion for a Hamiltonian's ground state to be sign-free (i.e.\ have no sign structure): that all off-diagonal matrix elements are non-positive \cite{tarazagaPerronFrobeniusTheorem2001}.
Hamiltonian matrices that satisfy this criterion are referred to as \textit{stoquastic} \cite{bravyiComplexityStoquasticLocal2006}, but Eq.~\eqref{eq:hamiltonian} in the computational basis does not generally fall into this class.
Thus, realizations of Eq.~\eqref{eq:hamiltonian} may have a ground state with sign structure.
However, we argue below that this need not necessarily be a barrier to state reconstruction with standard generative models.
In order to make this argument, we briefly explore some consequences of the sign structure in the context of quantum state reconstruction.

Given a normalized state $\ket{\psi}$ with real expansion coefficients $\ip{\vec{n}}{\psi}$ in some computational basis $\mathcal{B}$, we fix an arbitrary $\ket{\vec{n}^\star} \in \mathcal{B}$ and partition the set of all basis states,
\begin{align}
    \mathcal{B}
     & = \{ \ket{\vec{n}} \} = \mathcal{B}^{+} \cup \mathcal{B}^{-},
\end{align}
into the equivalence classes
\begin{subequations}
    \begin{align}
        \mathcal{B}^{+}
         & = \{ \ket{\vec{n}} \in \mathcal{B} \mid \ip{\vec{n}}{\psi} \ip{\vec{n}^\star}{\psi} \ge 0 \} \\
        \intertext{and}
        \mathcal{B}^{-}
         & = \{ \ket{\vec{n}} \in \mathcal{B} \mid \ip{\vec{n}}{\psi} \ip{\vec{n}^\star}{\psi} < 0 \}.
    \end{align}
\end{subequations}
The former contains those states $\ket{\vec{n}}$ whose coefficients $\ip{\vec{n}}{\psi}$ have the same sign as $\ip{\vec{n}^\star}{\psi}$ (by convention called ``positive'', disregarding any global phase), while the latter states have coefficients with the opposite sign (called ``negative'').
Indeed our formalism could be generalized to the case where the expansion coefficients are non-trivially complex, however we do not present that result in this paper.
We emphasize that in our ED and DMRG calculations of Eq.~\eqref{eq:hamiltonian}, we observe that all coefficients are real numbers.

We let
\begin{align}
    \tau^{\pm}
     & = \mel{\psi}{\hat{\mathcal{P}}^{\pm}}{\psi}
    = \smashoperator{\sum_{\ket{\vec{n}} \in \mathcal{B}^{\pm}}} \abs{\ip{\vec{n}}{\psi}}^2
\end{align}
be the total contributions to the wavefunction from positive ($\tau^{+}$) and negative ($\tau^{-}$) coefficients, where
\begin{align}
    \hat{\mathcal{P}}^{\pm}
     & = \smashoperator{\sum_{\ket{\vec{n}} \in \mathcal{B}^{\pm}}} \dyad{\vec{n}}
\end{align}
are projection operators satisfying $\hat{\mathcal{P}}^{+} + \hat{\mathcal{P}}^{-} = \hat{\mathds{1}}$.
The normalization condition demands $\tau^{+} + \tau^{-} = 1$.

We define the rectified state,
\begin{align}
    \ket*{\psi_{||}}
     & = \hat{\mathcal{P}}^{+} \ket{\psi} - \hat{\mathcal{P}}^{-} \ket{\psi},
\end{align}
which is normalized by construction, and whose coefficients may be written as
\begin{align}
    \ip*{\vec{n}}{\psi_{||}}
     & = \sgn\!\qty(\ip{\vec{n}^\star}{\psi}) \, \abs{\ip{\vec{n}}{\psi}},
\end{align}
making it clear that they are all positive.
When $\mathcal{B}^{-} = \varnothing$ and $\tau^{-} = 0$, the original state is sign-free (all coefficients are positive) and coincides with the rectified state: $\ket*{\psi_{||}} = \ket{\psi}$.
Otherwise, the rectified state is only an approximation to the original state, since all relative phase information is lost.
The overlap between these states is given by
\begin{align}
    \ip*{\psi_{||}}{\psi}
     & = \mel{\psi}{\left(\hat{\mathcal{P}}^{+} - \hat{\mathcal{P}}^{-}\right)}{\psi}
    = 1 - 2 \tau^{-},
\end{align}
so $\tau^{-}$ quantifies the amount of error introduced into the state by the approximation.
As shown in Appendix~\ref{sec:rectified-expval}, when $\ket{\psi}$ is an eigenstate of a real Hamiltonian $\hat{H}$ with energy $E$, the error in the evaluation of the energy arising due to the state approximation is
\begin{subequations}
    \begin{align}
        \varepsilon
         & = \mel*{\psi_{||}}{\hat{H}}{\psi_{||}} - \mel{\psi}{\hat{H}}{\psi}                         \\
         & = -4 E \tau^{-} + 4 \mel{\psi}{\hat{\mathcal{P}}^{-} \hat{H} \hat{\mathcal{P}}^{-}}{\psi}.
    \end{align}
\end{subequations}

\begin{figure}
    \begin{center}
        \includegraphics[width=\linewidth]{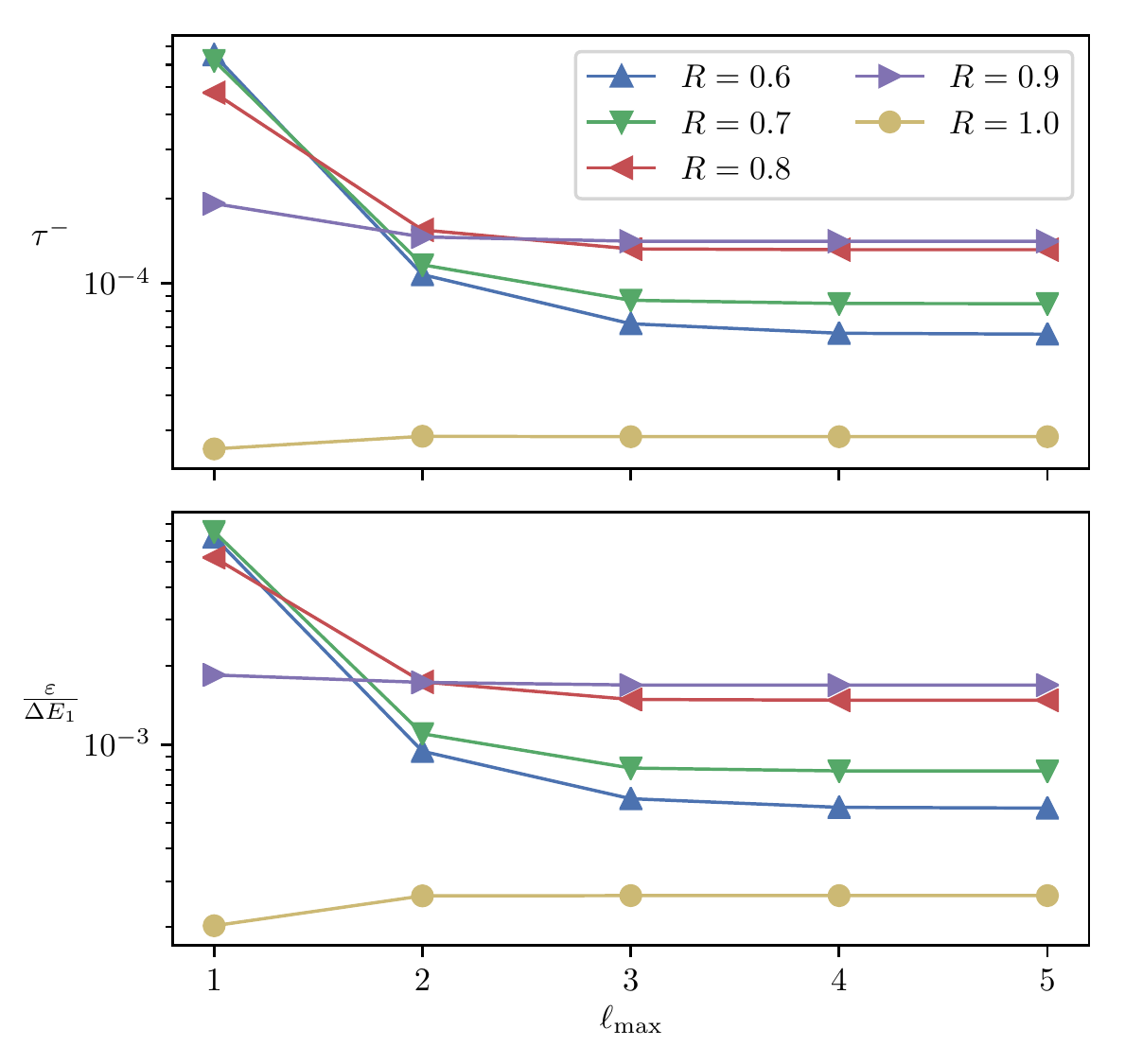}
    \end{center}
    \caption{
        Convergence of the rectified state error measures $\tau^{-}$ (top panel) and $\varepsilon / \Delta E_1$ (bottom panel) with basis size truncation parameter $\ell_\mathrm{max}$, for $N = 6$ and a variety of $R$ values, computed via exact diagonalization.
        On this scale, crude convergence is achieved by $\ell_\mathrm{max} = 3$, and there is no visible improvement from $\ell_\mathrm{max} = 4$ to $\ell_\mathrm{max} = 5$.
    }
    \label{fig:negcoef_convergence}
\end{figure}

\begin{figure}
    \begin{center}
        \includegraphics[width=\linewidth]{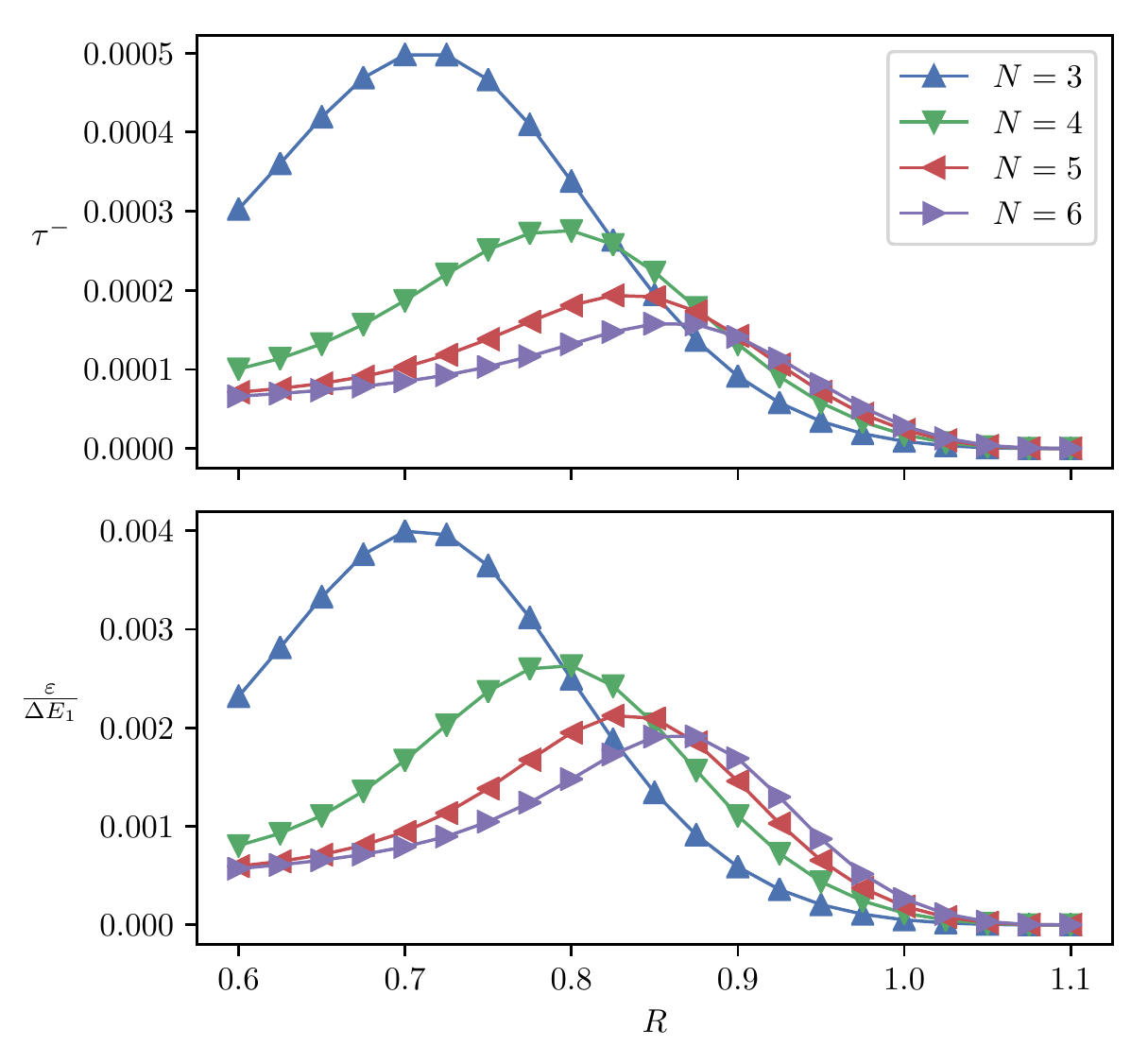}
    \end{center}
    \caption{
        The rectified state error measures $\tau^{-}$ (top panel) and $\varepsilon / \Delta E_1$ (bottom panel) for $\ell_\mathrm{max} = 5$, computed via exact diagonalization.
    }
    \label{fig:negcoef_values}
\end{figure}

We employ exact diagonalization to compute $\tau^{-}$ and $\varepsilon / \Delta E_1$ for the ground state of Eq.~\eqref{eq:hamiltonian}, using the basis truncation described in Ref.~\onlinecite{halversonQuantifyingEntanglementRotor2018}.
We set the energy scale to be $\Delta E_1$, which is the energy gap to the first excited state, and we take $\ket{\vec{n}^\star}$ to be the non-interacting rotor ground state, $\ket{0 \, 0, \ldots, 0 \, 0}$.
From Fig.~\ref{fig:negcoef_convergence}, it is clear that for the selected system parameters $R$ and $N$, both quantities converge to small positive values with increasing basis truncation parameter $\ell_\mathrm{max}$.
The converged values (at $\ell_\mathrm{max} = 5$) are shown for an assortment of $R$ and $N$ combinations in Fig.~\ref{fig:negcoef_values}.
Both measures of error exhibit a peak as a function of separation distance $R$, but this peak shifts and shrinks with increasing system size $N$.
Based on these results, we expect that for the parameters considered in this work ($R \ge 1$, $N \le 8$), $\tau^{-}$ will be less than $10^{-4}$, and $\varepsilon / \Delta E_1$ will not exceed 0.1\%.

In Sec.~\ref{sec:reconstruction}, we detail the reconstruction of the ground state of Eq.~\eqref{eq:hamiltonian} from projective measurement data in the free-rotor basis.
We consider training up to an energy threshold of 5\% (see Sec.~\ref{sec:training_evals}), which is substantially greater than $\varepsilon / \Delta E_1$,
hence justifying the use of only a single basis (Eq.~\eqref{eq:ref_basis}).

\section{Quantum state reconstruction}
\label{sec:reconstruction}

In this section, we develop a machine learning method for the reconstruction of the ground state of a linear chain of $N$ dipolar molecules as a model of endofullerene NMAs,
with the Hamiltonian given in Eq.~\eqref{eq:hamiltonian}.  The reconstruction proceeds by training a generative model, the restricted Boltzmann machine (RBM), on projective measurement data in the angular momentum basis only.  The use of this as a computational basis is justified in Sec.~\ref{sec:sign_structure}.
Quantum state reconstruction with RBMs
is explained in detail in many other works \cite{torlaiLearningThermodynamicsBoltzmann2016, torlaigiacomoAugmentingQuantumMechanics2018, hintonPracticalGuideTraining2012, salakhutdinovRestrictedBoltzmannMachines2007, murphyMachineLearningProbabilistic2012}.
However, a brief explanation of the expanded RBM scheme used in this study is offered here.

\subsection{Multinomial Restricted Boltzmann Machine} \label{sec:RBM_theory}

The RBM is a two-layer, bidirectionally-connected neural network whose goal is to reconstruct a target probability distribution $q(\bm{\sigma})$ from data $\bm{\sigma}$ sampled from this distribution.
In our case, the target distribution is taken to be the square of the ground state of Eq.~\eqref{eq:hamiltonian},
\begin{equation} \label{eq:target_dist}
    q(\bm{\sigma}) = \abs{\psi_{\mathrm{Exact}}(\bm{\sigma})}^2
    = \abs{\ip{\bm{\sigma}}{\psi_{\mathrm{Exact}}}}^2.
\end{equation}
We can calculate this ground state wavefunction using exact diagonalization (ED) or DMRG \cite{iouchtchenkoGroundStatesLinear2018}.
The data sets consist of one-hotted projective measurements on each of the $N$ sites in the reference basis (Eq.~\eqref{eq:ref_basis}) $\sigma_{id}$ where $i$ indexes the $N$ sites and $d$ indexes the $\mathcal{D}$ possible projective measurement outcomes (Eq.~\eqref{eq:local_hilbert}). In other words, $\sigma_{id} = 1$ corresponds to rotor $i$ having $(\ell_i,m_i) = \sigma_i = d$.
Such a data set can be generated for small system sizes by ED.  For larger system sizes,
Ref.~\onlinecite{iouchtchenkoGroundStatesLinear2018} presented a DMRG framework for calculating the ground state of Eq.~\eqref{eq:hamiltonian},
and the sampling algorithm by Ferris and Vidal can be used to generate training data in the $\ket{\ell m}$ basis\cite{ferrisPerfectSamplingUnitary2012}.

Outside of machine learning in the physical sciences, multinomial data is prevalent. For instance, a dictionary has its elements (words) mapped onto a one-hotted vector space in order to facilitate natural language processing.
Another example is collaborative filtering, for which an extension of the traditional binary RBM to multinomial data was introduced
as an application to the Netflix Prize in Ref.~\onlinecite{salakhutdinovRestrictedBoltzmannMachines2007}.
Such multinomial RBMs are highly suitable as generative models for the purpose of quantum state reconstruction, for example when the input datasets
contain multiple outcomes, like POVMs \cite{carrasquillaReconstructingQuantumStates2018}.
In the following, we adapt a multinomial RBM for use on the ground state of Eq.~\eqref{eq:hamiltonian}
with data in the free rotor basis with a local Hilbert space size given by Eq.~\eqref{eq:local_hilbert}.

\begin{figure}
    \begin{center}
        \includegraphics[width=\linewidth]{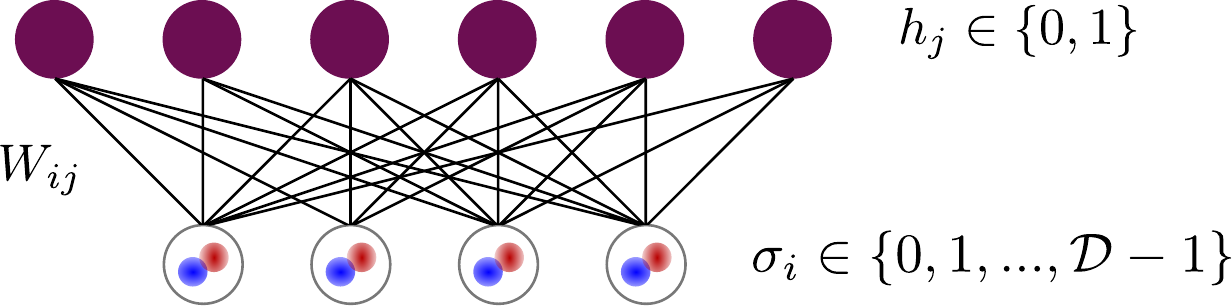}
    \end{center}
    \caption{Graphical representation of a multinomial RBM for learning rotor states.
        The visible nodes $\bm{\sigma}$ and the hidden nodes $\bm{h}$ are represented as endofullerene cartoons and purple nodes, respectively.
        Connecting each visible unit to each hidden unit is a weight tensor, $\bm{W}$.
        The external biases that couple to the visible ($\bm{b}$) and hidden ($\bm{c}$) layers are omitted from this depiction.}
    \label{fig:neuralnet}
\end{figure}

All $N$ sites form the \textit{visible} layer of the network, while interdependencies between sites in the visible layer are captured through $n_h$ binary \textit{hidden} nodes, taking on the values $h_j$ (see Fig.~\ref{fig:neuralnet}).
Each site in the visible layer is connected to each node in the hidden layer through a set of weights, $\bm{W}$, and each site and hidden node is coupled to an external bias field, $\bm{b}$ and $\bm{c}$ respectively; we refer to all the parameters simultaneously as $\bm{\lambda}$ = $(\bm{W}, \bm{b}, \bm{c})$.
There are no intralayer connections.
At the core of this model is a joint probability distribution over both layers defined by a Boltzmann distribution,
\begin{equation} \label{eq:full_prob}
    p_{\bm{\lambda}}\left(\bm{\sigma},\bm{h}\right)= \frac{e^{-E_{\bm{\lambda}}\left(\bm{\sigma},\bm{h}\right)}}{Z_{\bm{\lambda}}},
\end{equation}
with an energy
\begin{equation} \label{eq:energy}
    \begin{aligned}
        E_{\bm{\lambda}}\left(\bm{\sigma},\bm{h}\right) = & -\sum_{i=1}^{N} \sum_{j=1}^{n_{\mathrm{h}}} \sum_{d=1}^{\mathcal{D}} W_{ijd} h_j \sigma_{id} \\ &- \sum_{i=1}^{N} \sum_{d=1}^{\mathcal{D}} \sigma_{id} b_{id} \\ &- \sum_{j=1}^{n_{\mathrm{h}}} h_j c_j,
    \end{aligned}
\end{equation}
and a partition function (normalization)
\begin{equation} \label{eq:partition}
    Z_{\bm{\lambda}} = \sum_{\bm{\sigma},\bm{h}} e^{-E_{\bm{\lambda}}\left(\bm{\sigma},\bm{h}\right)}.
\end{equation}

The marginal distribution over the visible layer, $p_{\bm{\lambda}}\left(\bm{\sigma}\right)$, can be obtained by summing over the hidden layer:
\begin{equation} \label{eq:marginal_prob}
    p_{\bm{\lambda}}\left(\bm{\sigma}\right) = \sum_{\bm{h}} p_{\bm{\lambda}}\left(\bm{\sigma},\bm{h}\right) = \frac{e^{-\mathcal{E}_{\bm{\lambda}}\left(\bm{\sigma}\right)}}{Z_{\bm{\lambda}}},
\end{equation}
where,
\begin{equation} \label{eq:eff_energy}
    \begin{aligned}
        \mathcal{E}_{\bm{\lambda}}\left(\bm{\sigma}\right) = & - \sum_{i=1}^{N} \sum_{d=1}^{\mathcal{D}} b_{id} \sigma_{id} \\ &- \sum_{j=1}^{n_{\mathrm{h}}} \ln(1+\mathrm{exp}\left(c_j + \sum_{i=1}^{N} \sum_{d=1}^{\mathcal{D}}W_{ijd} \sigma_{id} \right) ).
    \end{aligned}
\end{equation}
With the multinomial RBM thus defined, training proceeds as usual through minimization of the Kullback-Leibler divergence.  Many of the details of training
of the multinomial RBM are similar to a conventional RBM with binary hidden and visible layers.
We provide these remaining details in Appendix~\ref{sec:TrainingRBM} for completeness.

\subsection{Training Evaluators} \label{sec:training_evals}

We are interested in training an RBM from synthetic data generated from the exact ground state of Eq.~\eqref{eq:hamiltonian}, obtained from ED or DMRG.
A central question is the quality of the quantum state that is reconstructed by the RBM.  Here, we develop a criterion to assess the quality of reconstruction, which can be
used on endofullerene systems of various sizes $N$.  For example, one common criterion is the fidelity between the RBM state and the exact ground state.
However, the fidelity is expensive to compute for large $N$, since the full marginal distribution $p_{\bm{\lambda}}$ must be calculated, including the normalization $Z_{\bm{\lambda}}$.
As noted previously\cite{sehayekLearnabilityScalingQuantum2019}, it is much more efficient to take advantage of the simplicity of Gibbs sampling in the RBM, and to calculate the expectation value of a local observable such as the energy.

In order to treat the RBM as a wavefunction ansatz, we must take the square root of the marginal distribution:
\begin{align}
    \psi_{\bm{\lambda}}(\bm{\sigma})
     & = \sqrt{p_{\bm{\lambda}}(\bm{\sigma})}.
\end{align}
The probability values are strictly non-negative, so the resulting state $\ket{\psi_{\bm{\lambda}}}$ is inherently real and sign-free in this basis.
When the RBM is sufficiently well trained (i.e.~$p_{\bm{\lambda}}(\bm{\sigma}) \approx q(\bm{\sigma})$), the state that it encodes should be the rectified version of $\ket{\psi_{\mathrm{Exact}}}$.
It is then straightforward (at least in principle) to compute the RBM expectation value of an arbitrary operator $\hat{A}$ as
\begin{align}
    \langle \hat{A} \rangle_{\psi_{\bm{\lambda}}}
     & = \mel{\psi_{\bm{\lambda}}}{\hat{A}}{\psi_{\bm{\lambda}}}
    = \sum_{\bm{\sigma}\bm{\sigma}'} \psi_{\bm{\lambda}}\left(\bm{\sigma}\right) \psi_{\bm{\lambda}}\left(\bm{\sigma}'\right) A_{\bm{\sigma} \bm{\sigma}'}, \label{eq:MC_general}
\end{align}
where $A_{\bm{\sigma} \bm{\sigma}'} = \mel{\bm{\sigma}}{\hat{A}}{\bm{\sigma}'}$.
In practice, this explicit summation is not tractable, so it is replaced by Monte Carlo sampling.

To evaluate the expectation value of Eq.~\eqref{eq:hamiltonian}, it is useful to separate it into its diagonal and off-diagonal terms;
i.e.~the rotational energy, $\hat{K}$, and dipole-dipole potential, $\hat{V}$, respectively.
Given a set of samples, $\mathbf{G} = \{ \bm{\sigma}^g \}$, produced from $p_{\bm{\lambda}}$ by Gibbs sampling, one can trivially approximate the expectation value of the rotational energy term as a diagonal Monte Carlo estimator:
\begin{subequations}
    \begin{align}
        \langle \hat{K} \rangle_{\psi_{\bm{\lambda}}} = & \sum_{\bm{\sigma}} \vert  \psi_{\bm{\lambda}}\left(\bm{\sigma}\right) \vert^2 \ell_{\bm{\sigma}}^2 \approx \frac{1}{\vert \mathbf{G} \vert} \sum_{g=1}^{\vert \mathbf{G} \vert} \ell_{\bm{\sigma}^g}^2, \label{eq:MC_kinetic} \\
        \intertext{where}
        \ell_{\bm{\sigma}}^2 =                          & \mel{\bm{\sigma}}{\hat{K}}{\bm{\sigma}} = \sum_{i=1}^N \ell_{i}(\ell_i + 1).
    \end{align}
\end{subequations}
To avoid calculating $\psi_{\bm{\lambda}}(\bm{\sigma})$ for the potential energy term, the general expression in Eq.~\eqref{eq:MC_general} can be approximated as an off-diagonal Monte Carlo estimator\cite{beachQuCumberWavefunctionReconstruction2019, Microsoft_Tomo}:
\begin{equation} \label{eq:MC_potential}
    \langle \hat{V} \rangle_{\psi_{\bm{\lambda}}} \approx \frac{1}{\vert \mathbf{G} \vert} \sum_{g=1}^{\vert \mathbf{G} \vert} \sum_{\bm{\sigma}'} \frac{ \tilde{\psi}_{\bm{\lambda}}\left(\bm{\sigma}'\right)}{\tilde{\psi}_{\bm{\lambda}}\left(\bm{\sigma}^{g}\right)} V_{\bm{\sigma}^{g} \bm{\sigma}'},
\end{equation}
where
\begin{align}
    \tilde{\psi}_{\bm{\lambda}}(\bm{\sigma})
     & = \psi_{\bm{\lambda}}(\bm{\sigma}) \sqrt{Z_{\bm{\lambda}}}
    = e^{-\mathcal{E}(\bm{\sigma}) / 2}
\end{align}
is the unnormalized RBM wavefunction, which may be evaluated without knowledge of the normalization $Z_{\bm{\lambda}}$.
As $V_{\bm{\sigma} \bm{\sigma}'}$ is very sparse \cite{iouchtchenkoGroundStatesLinear2018}, most of the terms in the sum over $\bm{\sigma}'$ in Eq.~\eqref{eq:MC_potential} may be skipped, avoiding the unfavorable scaling of the basis size.
The total energy can therefore be efficiently approximated by adding Eqs.~\eqref{eq:MC_kinetic} and \eqref{eq:MC_potential}:
\begin{equation} \label{eq:E_RBM}
    E_{\mathrm{RBM}}
    \approx \frac{1}{\vert \mathbf{G} \vert} \sum_{g=1}^{\vert \mathbf{G} \vert} \ell_{\bm{\sigma}^g}^2 + \frac{1}{R^3} \sum_{\bm{\sigma}'} \frac{ \tilde{\psi}_{\bm{\lambda}}\left(\bm{\sigma}'\right)}{\tilde{\psi}_{\bm{\lambda}}\left(\bm{\sigma}^{g}\right)} V_{\bm{\sigma}^{g} \bm{\sigma}'}.
\end{equation}

Then, the training of the RBM can be evaluated by periodically generating samples from the RBM, and calculating the energy from those samples, along with the relative error in this energy as compared to the exact value.
As explained in Appendix~\ref{sec:threshold}, we use
\begin{equation} \label{eq:delta}
    \delta = \abs{\frac{E_{\mathrm{RBM}} - E_{\mathrm{Exact}}}{\Delta E_1}}
\end{equation}
as a measure of the quality of the state encoded in the RBM, with the exact energy and excitation gap calculated by ED or DMRG.
In Sec.~\ref{sec:results}, we will use $\delta = 5\%$ as a {\it learning criterion}, in analogy to that defined in the scaling studies of Ref.~\onlinecite{sehayekLearnabilityScalingQuantum2019}.

\subsection{Results} \label{sec:results}

\begin{figure}
    \begin{center}
        \includegraphics[width=\linewidth]{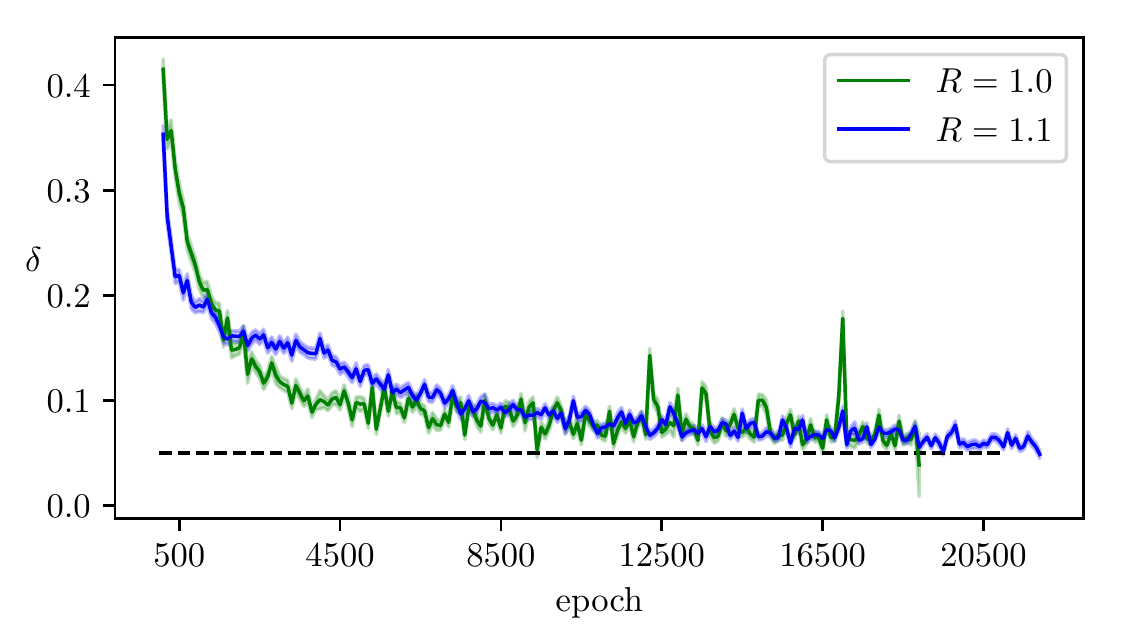}
    \end{center}
    \caption{The error in the energy, $\delta$, during the training process for $N=4$ and $R=1.0, 1.1$. For $R=1.0, 1.1$, we used $n_h = 5, 4$, respectively, along with 10,000 input samples. The shaded region around each line is the standard error in $\delta$.}
    \label{fig:training}
\end{figure}

Endofullerene NMAs defined by Eq.~\eqref{eq:hamiltonian}
are characterized solely by the parameter $R$, which determines the relative strengths of the kinetic and potential energy terms.
Previous Monte Carlo, path integral ground state, and DMRG simulations on finite-size systems have provided evidence that a continuous phase transition exists near $R=1$ \cite{iouchtchenkoGroundStatesLinear2018,abolinsGroundStateMonte2011, sahooPathIntegralGround2020}.
We refer to this as the {\it critical} value of $R$, for simplicity.
In the following, we consider two values of $R$: $R = 1.0$ and $1.1$, both with $\ell_{\mathrm{max}} = 3$.
For all RBMs, $\delta = 0.05$ was used as the learning criterion, i.e.~the threshold at which training was considered to be complete.
Note that since the endofullerene NMAs that we study are for finite sizes $N$, systems at criticality will not display gap-closing.
To calculate the energy from samples generated by the RBM during training ($E_\mathrm{RBM}$ from Eq.~\eqref{eq:E_RBM}), 7500 to 10,000 independent samples were generated using 7500 to 10,000 Gibbs steps for $N = $ 2, 4, 6 and 8. An example of the training process for $N=4$ is given in Fig.~\ref{fig:training}.

The hidden layer serves the purpose of acting as a feature detector for the visible layer of the RBM.
It is therefore expected that for systems that have long correlation lengths or high entanglement, a large number of hidden units may be
required to obtain an accurate reconstruction.
Similarly, quantum states with a larger number of non-trivial amplitudes in the computational basis will require more training data (i.e.~higher sample complexity)
in order to obtain an accurate reconstruction.
In the language of one-dimensional chains of dipolar rotors governed by Eq.~\eqref{eq:hamiltonian}, one expects
$R$ = 1.1 to possess fewer non-trivial configurations outside of the free-rotor ground state ($\ket{0 \, 0, \ldots, 0 \, 0}$)
than at the critical value of $R$ = 1.0.
The relative contribution of the free-rotor ground state as a function of $N$ and $R$ is given in Tab.~\ref{tab:wf_ratio}.
Evidently, for larger $R$ and smaller $N$, the free-rotor ground state's probability amplitude vastly outweighs that of all other probability amplitudes.
\begin{table}
    \centering
    \caption{The ratio between the free-rotor ground state probability amplitude and the sum of all other probability amplitudes, $\abs{\psi_{\rm Exact}(0)}^2 / \sum_{i\neq0}\abs{\psi_{\rm Exact}(i)}^2$, as calculated via DMRG.}
    \begin{tabular}{lccc}
        \toprule
                & $N=2$ & $N=3$ & $N=4$ \\
        \hline
        $R=1.0$ & 24.7  & 10.3  & 5.9   \\
        $R=1.1$ & 274.1 & 132.2 & 86.6  \\
        \botrule
    \end{tabular}  \label{tab:wf_ratio}
\end{table}

\begin{figure} [t]
    \begin{center}
        \includegraphics[width=\linewidth]{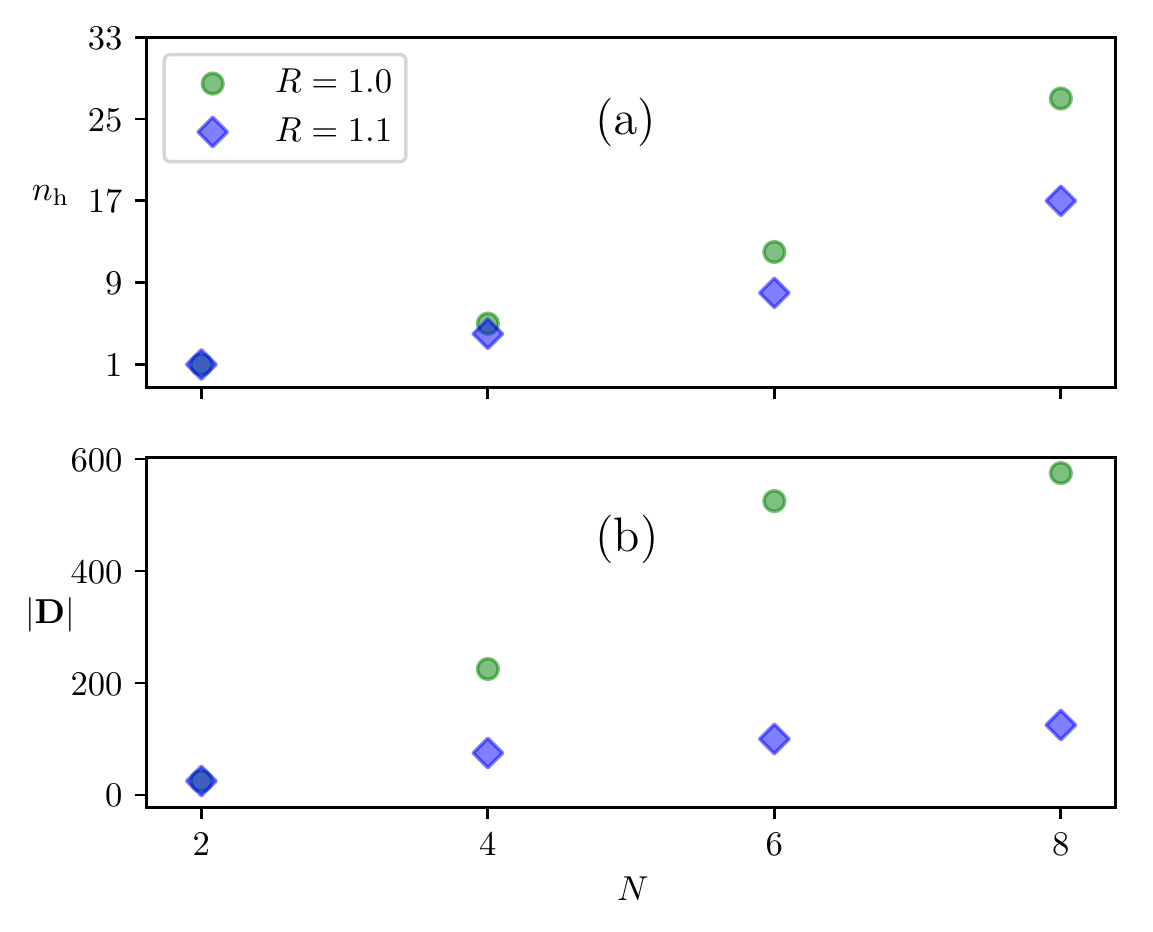}
    \end{center}
    \caption{ (a) The minimum hidden layer size ($n_{h,\mathrm{min}}$) required to train an RBM to $\delta=5\%$ given a saturated input data size. (b) The minimum input data size ($\abs{\mathbf{D}}_{\mathrm{min}}$) required to train an RBM to $\delta=5\%$ given a saturated hidden layer size.}
    \label{fig:scaling}
\end{figure}

From the above arguments, we expect that the ground state near the critical point ($R=1.0$) should be
harder to learn, both in terms of hidden layer size and sample complexity, than off-critical systems (such as $R=1.1$).
This hypothesis can be tested as a function of $N$, which can be used to elucidate the scaling behavior as in Ref.~\onlinecite{sehayekLearnabilityScalingQuantum2019}.
To do so, we train a variety of system sizes $N$, systematically increasing resources (hidden units, or data set size) until the RBM energy satisfies our learning criterion $\delta = 5\%$.
This defines the minimum resources required for the reconstruction.
The results of this procedure are shown in Fig.~\ref{fig:scaling}.  Plot (a) shows the scaling of the ground state reconstruction with minimum hidden layer size ($n_h$).
Similar to the procedure defined in Ref.~\onlinecite{sehayekLearnabilityScalingQuantum2019}, to make this scaling plot we assume access to unlimited data.
In practice we find that $\abs{\mathbf{D}} = \order{10^4}$ completely saturates the training data sample complexity for the system sizes studied.
Fig.~\ref{fig:scaling}(a) empirically defines $n_{h,\mathrm{min}}$ as a function of $N$.
As evident from the plot, not only does the critical point $R=1.0$ require a larger number of hidden units than the off-critical $R=1.1$ for system sizes $N>2$, the functional
form of the scaling curve also appears to be different.
Next, in Fig.~\ref{fig:scaling}(b), we fix the number of hidden units in the RBM to $n_{h,\mathrm{min}} + 1$, and systematically increase the size of the training data set
$\abs{\mathbf{D}}$ until the learning criterion is satisfied.
As hypothesized, the critical value of $R$ requires a larger sample complexity for every $N>2$, although it is harder to discern a simple scaling curve for $R=1.0$.
In both cases, one may question whether these scaling results lie in the asymptotic limit; clearly, future work on larger system sizes will be required to answer this.

Finally, we comment on efficiency issues inherent to the sampling procedure of the RBM, called {\it block Gibbs} sampling (see Appendix~\ref{sec:TrainingRBM}).
In any procedure that uses Gibbs sampling, particularly training or estimator calculation, the state of each visible node is produced independently from the others based on a
conditional probability involving the state of all hidden nodes.  The ``restricted'' nature of the RBM ensures this independence, which also means that conditional relationships
between different visible units cannot be directly enforced.  In other words, it is difficult to implement symmetry restrictions on visible states in a standard RBM that manifest as null probability amplitudes in the true ground state.
Recall, as discussed in Section \ref{sec:Ham_Hilb}, the Hamiltonian in Eq.~\eqref{eq:hamiltonian} conserves the total $m$ value $m = \sum_{i=1}^N m_i$ and the total $\ell$ parity $\ell_\mathrm{p} \equiv \sum_{i=1}^N \ell_i \pmod{2}$.
In the free-rotor basis (Eq.~\eqref{eq:ref_basis}), this manifests in the ground state wavefunction as null probability amplitudes for configurations that violate these symmetries.

\begin{figure} [ht!]
    \begin{center}
        \includegraphics[width=\linewidth]{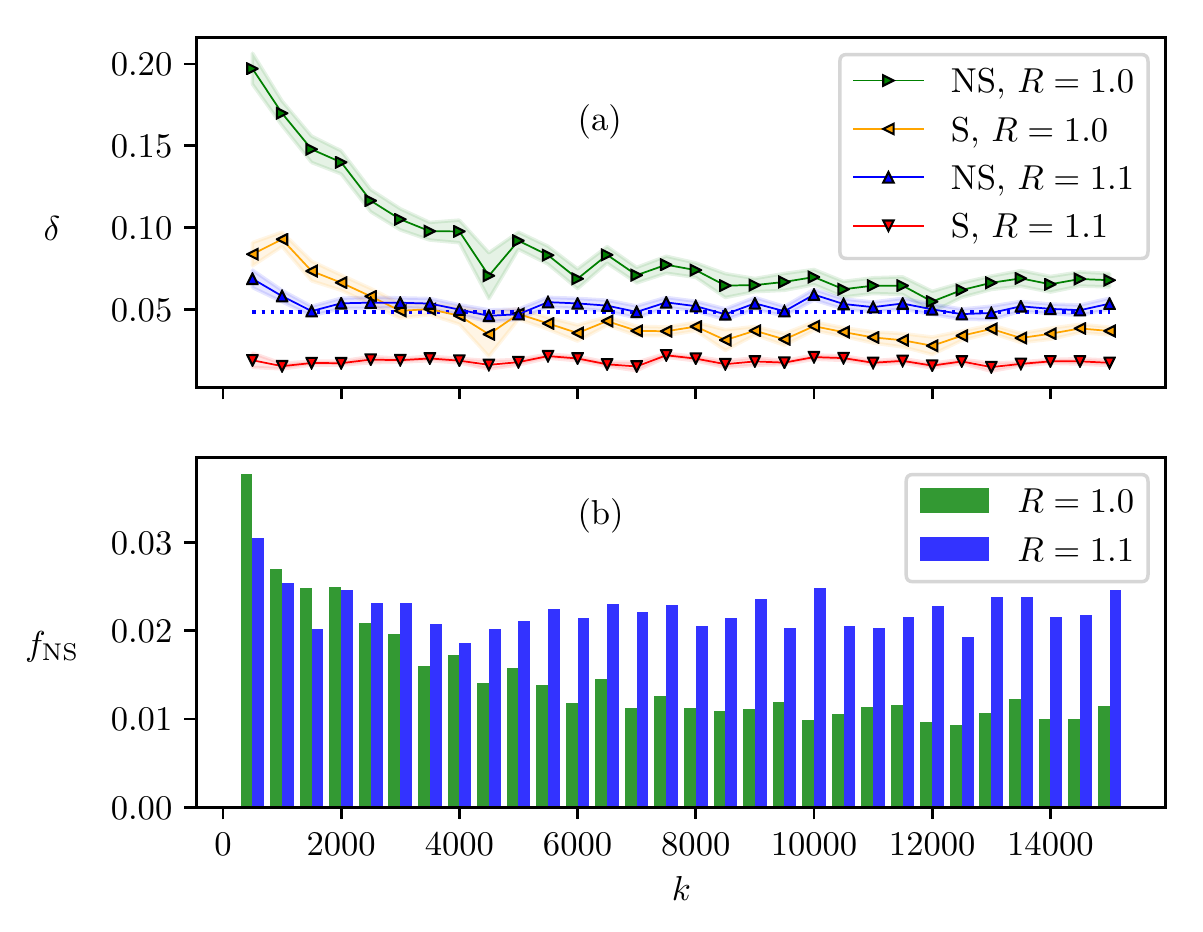}
    \end{center}
    \caption{(a) $\delta$ as a function of the number of Gibbs sampling steps, $k$, for trained RBMs for $N=6$ using $n_{h,\mathrm{min}}$. Data denoted by $\mathrm{S}, R = 1.0,1.1$ correspond to $\delta$ as calculated by only counting samples satisfying the symmetries of Eq.~\eqref{eq:hamiltonian}, while $\mathrm{NS}, R = 1.0,1.1$ included all samples generated. The dashed line represents the value of $\delta$ that the RBMs are trained to. At each Gibbs sampling step, 10,000 independent samples were generated to calculate $\delta$. (b) The fraction of samples, $f_{\mathrm{NS}}$, generated at a given number of Gibbs sampling steps, $k$, that violate the symmetries of Eq.~\eqref{eq:hamiltonian}.}
    \label{fig:equilibration}
\end{figure}

Fig.~\ref{fig:equilibration}(a) shows $\delta$ (Eq.~\eqref{eq:delta}) as a function of the number of Gibbs steps $k$ as calculated from trained RBMs for $N = 6$ (when initialized to an all-zero configuration).
Even for our threshold of $\delta \leq 5\%$, a non-negligible amount of samples generated from the RBM violate the total $m$ and $\ell_\mathrm{p}$ symmetries as evident in Fig.~\ref{fig:equilibration}(b).
If these symmetry-violating samples are removed, Fig.~\ref{fig:equilibration}(a) shows that the value of $\delta$ improves significantly, as expected.
We note that the Gibbs sampling chain equilibration time here is much larger than that required for similar studies on the transverse-field Ising model\cite{sehayekLearnabilityScalingQuantum2019}.
Such shortcomings in generating symmetry-violating samples via Gibbs sampling can be overcome by other classes of generative models, in particular wavefunctions such as recurrent neural networks, where recent work has demonstrated the enforcement of symmetries in the autoregressive architecture \cite{hibat-allahRecurrentNeuralNetwork2020a}.

\section{Discussion} \label{sec:discussion}

In this paper we study NMAs of endofullerenes through a model of one-dimensional quantum rotors with Coulombic dipole-dipole interactions.
As part of an exploration of their suitability for quantum information processing purposes, we ask whether efficient quantum state reconstruction is possible for the
ground states of such systems, using recently developed machine learning techniques.
In the case of the dipolar quantum rotor chains, the model Hamiltonian is non-stoquastic in the angular momentum basis,
implying that a trivial (positive) sign structure is not guaranteed for the ground state wavefunction.
This fact holds significance for the prospects of reconstruction with generative models, which are more efficient for real and positive wavefunctions, with measurement outcomes
that can be mapped directly to classical probability distributions.
We study the ground state wavefunction numerically on finite system sizes using ED and DMRG.
By analyzing the relative error associated with neglecting the negative wavefunction amplitudes, we determine that their contribution is negligible in the
expectation value of the energy up to typical uncertainties in the machine learning reconstruction procedure.

In light of this, we demonstrate here a machine learning reconstruction technique of using an RBM with multinomial
visible units to capture the local Hilbert space of the dipolar rotor.  Using the free-rotor angular momentum states as a computational basis, we generate synthetic measurement
data using DMRG, and train a multinomial RBM on a variety of one-dimensional chains of length $N$.  Examining the energy difference between the RBM
and the DMRG, we find that typical accuracies obtained by the machine learning procedure are well above the errors associated with neglecting negative amplitudes.
Defining a {\it learning criterion} as 5\% difference, in units of the first excitation gap, we examine the scaling of the reconstruction for two values parameterizing
the strength of the dipolar interaction; one near a quantum critical point, and one farther away.
We observe that for the critical system, the size of the RBM hidden layer required to achieve the learning criterion
is larger than for the off-critical system, assuming the availability of a sufficiently large measurement dataset.
Conversely, holding the number of hidden units fixed near the minimum required to achieve the learning criterion, critical dipolar chains require significantly more data to train.

In the future, it would be interesting to continue this investigation for much larger system sizes to attempt to extract asymptotic scaling functions, as was done previously
at the critical point of the transverse-field Ising model\cite{sehayekLearnabilityScalingQuantum2019}.
Scaling functions could be investigated for multiple criteria on different local and non-local estimators, such as correlation functions or entanglement entropies.
From the observation in this work of significant fluctuations in the energy estimator of trained RBMs, it is possible that
other types of generative models that do not rely on Gibbs sampling may be more suitable for such investigations.
In particular, autoregressive models such as recurrent neural networks allow various symmetries to be taken into account when generating samples\cite{hibat-allahRecurrentNeuralNetwork2020a}.
Their use in quantum state reconstruction is being explored in earnest.

Finally, we note that there are in general two classes of quantum wavefunctions from the perspective of machine learning state reconstruction.
First are those that can be efficiently learned with standard generative models, which are designed to represent classical probability distributions.
This class requires measurement data in only the computational basis.
Second are those wavefunctions that do not have strictly real and positive amplitudes in the computational basis.  These require generative models that are
modified to account for the sign or phase structure, as well as requiring measurement data in a significant number of different bases to satisfy informational completeness.
However, as we have shown in the case of dipolar quantum rotors chains, some quantum systems that do not rigorously fall in the first class
may have a sufficiently small number of non-positive amplitudes to allow for an accurate reconstruction when one is interested in reproducing local observables.
It is an interesting theoretical question to ask how many physical systems fall into the class of states that are efficient to reconstruct
with standard machine learning techniques, in spite of their sign structure.

\subsection*{Acknowledgements}

We thank J. Carrasquilla, A. Golubeva, and G. Torlai for many enlightening discussions.
RGM is supported by the Natural Sciences and Engineering Research Council of Canada (NSERC),
Canada Research Chair (CRC) program, and the Perimeter Institute for Theoretical Physics.
Research at Perimeter Institute is supported in part by the Government of Canada through the Department of Innovation,
Science and Economic Development Canada and by the Province of Ontario through the Ministry of Colleges and Universities.
P.-N.R. (RGPIN-2016-04403) acknowledges NSERC, the Ontario Ministry of Research and Innovation (MRI), the CRC program (950-231024), the Canada Foundation for Innovation (CFI) (project No. 35232), and the Canada First Research Excellence Fund (CFREF).

\appendix

\section{Rectified State Expectation Values}
\label{sec:rectified-expval}

We wish to determine the error
\begin{align}
    \varepsilon
     & = \mel*{\psi_{||}}{\hat{A}}{\psi_{||}} - \mel{\psi}{\hat{A}}{\psi}
\end{align}
made in the expectation value of an operator $\hat{A}$ by using the rectified state $\ket*{\psi_{||}}$ as an approximation for $\ket{\psi}$.
By making use of the identity $\hat{\mathcal{P}}^{+} + \hat{\mathcal{P}}^{-} = \hat{\mathds{1}}$, the first term may be expressed as
\begin{align}
    \mel*{\psi_{||}}{\hat{A}}{\psi_{||}}
     & = \mel*{\psi}{\hat{A}}{\psi}
    - 2 \mel*{\psi}{\mathcal{P}^{+} \hat{A} \mathcal{P}^{-} + \mathcal{P}^{-} \hat{A} \mathcal{P}^{+}}{\psi},
\end{align}
so we immediately obtain the general expression
\begin{subequations}
    \begin{align}
        \varepsilon
         & = -2 \mel{\psi}{
            \hat{\mathcal{P}}^{+} \hat{A} \hat{\mathcal{P}}^{-}
            + \hat{\mathcal{P}}^{-} \hat{A} \hat{\mathcal{P}}^{+}
        }{\psi}                  \\
        \intertext{or, equivalently,}
        \varepsilon
         & = 2 \mel*{\psi_{||}}{
            \hat{\mathcal{P}}^{+} \hat{A} \hat{\mathcal{P}}^{-}
            + \hat{\mathcal{P}}^{-} \hat{A} \hat{\mathcal{P}}^{+}
        }{\psi_{||}}.
    \end{align}
\end{subequations}
The second form follows from $\hat{\mathcal{P}}^{\pm} \ket*{\psi_{||}} = \pm \hat{\mathcal{P}}^{\pm} \ket{\psi}$.
When $\hat{A}$ is Hermitian ($\hat{A}^\dagger = \hat{A}$), this simplifies to
\begin{align}
    \varepsilon
     & = -4 \Re \mel{\psi}{
        \hat{\mathcal{P}}^{+} \hat{A} \hat{\mathcal{P}}^{-}
    }{\psi},
\end{align}
while for diagonal $\hat{A}$ we get
\begin{align}
    \varepsilon & = 0.
\end{align}

\begin{figure}
    \begin{center}
        \includegraphics[width=\linewidth]{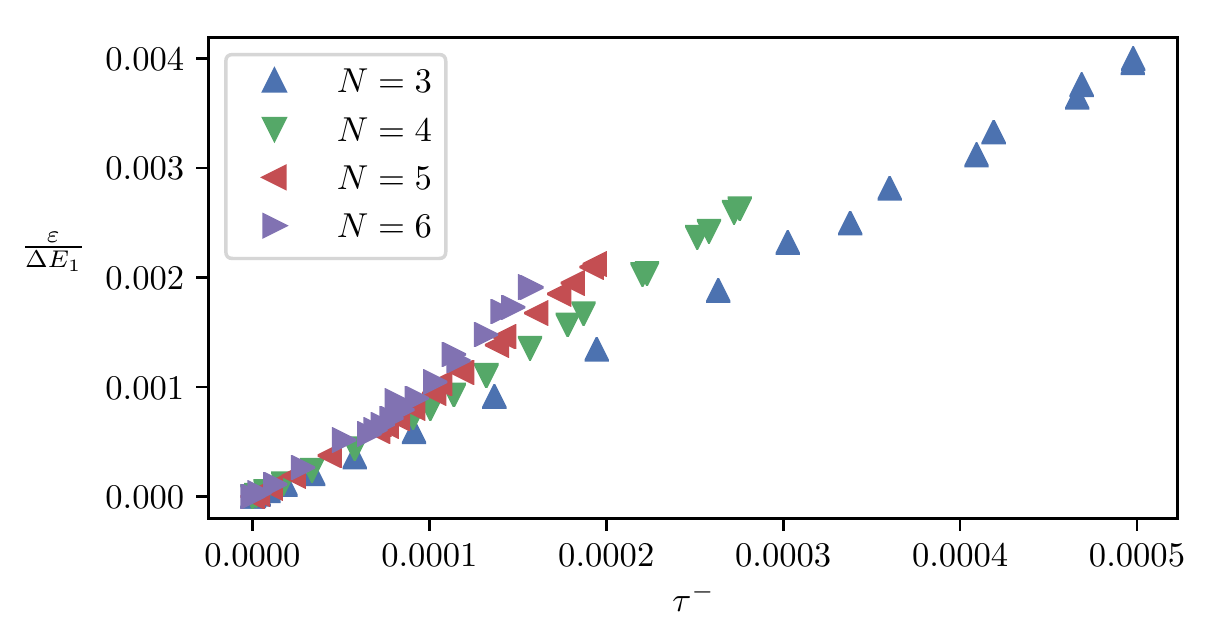}
    \end{center}
    \caption{
        Correlation between the rectification error measures in the top and bottom panels of Fig.~\ref{fig:negcoef_values}.
        Multiple points for a single system size $N$ correspond to different separation distances $R$.
    }
    \label{fig:negcoef_correlation}
\end{figure}

Of particular interest is the case of a real Hamiltonian $\hat{H}$ having $\ket{\psi}$ as an eigenstate with eigenvalue $E$.
The energy error due to the coefficient rectification is
\begin{align}
    \varepsilon
     & = -4 E \tau^{-} + 4 \mel{\psi}{\hat{\mathcal{P}}^{-} \hat{H} \hat{\mathcal{P}}^{-}}{\psi},
\end{align}
which suggests that $\tau^{-}$ and $\varepsilon$ are intimately connected.
This relationship is demonstrated for small systems in Fig.~\ref{fig:negcoef_correlation}, where the comparison of $\tau^{-}$ and $\varepsilon / \Delta E_1$ values from Fig.~\ref{fig:negcoef_values} across several combinations of the $R$ and $N$ system parameters reveals a positive correlation.

\section{Training the Restricted Boltzmann Machine} \label{sec:TrainingRBM}

To train the RBM (i.e.~tune the parameters, \bm{$\lambda$}, such that $p_{\bm{\lambda}}\left(\bm{\sigma}\right)$ approximates Eq.~\eqref{eq:target_dist}), the \textit{Kullback-Leibler (KL) divergence} is used as the RBM's cost function, $C_{\bm{\lambda}}$, where
\begin{equation} \label{eq:KL}
    C_{\bm{\lambda}} = \sum_{\bm{\sigma}} q\left(\bm{\sigma}\right) \ln(\frac{q\left(\bm{\sigma}\right)}{p_{\bm{\lambda}}\left(\bm{\sigma}\right)}).
\end{equation}
The optimization is carried out through stochastic gradient decent with the update rule
\begin{equation}
    \bm{\lambda} \leftarrow \bm{\lambda} - \eta \nabla_{\bm{\lambda}} C_{\bm{\lambda}},
\end{equation}
\begin{equation}
    \nabla_{\bm{\lambda}} = \left[ \frac{\partial}{\partial W_{ijd}} , \frac{\partial}{\partial b_{id}} , \frac{\partial}{\partial c_j} \right],
\end{equation}
where $\eta$ is referred to as the \textit{learning rate} and $\nabla_{\bm{\lambda}} C_{\bm{\lambda}}$ = $-\nabla_{\bm{\lambda}} \left\langle \ln(p_{\bm{\lambda}}\left(\bm{\sigma}\right))\right\rangle_q$.
The expectation value over the target distribution $q\left(\bm{\sigma}\right)$ is approximated to be the empirical distribution of a set of training data samples, $\mathbf{D}$, (many sets of local projective measurements of $\ell$ and $m$ for each site).
$\nabla_{\bm{\lambda}} C_{\bm{\lambda}}$ can then be simplified to
\begin{equation} \label{eq:grad_update}
    \nabla_{\bm{\lambda}} C_{\bm{\lambda}} \approx \left\langle \nabla_{\bm{\lambda}} \mathcal{E}_{\bm{\lambda}}\left(\bm{\sigma}\right) \right\rangle_{\mathbf{D}} - \left\langle \nabla_{\bm{\lambda}} \mathcal{E}_{\bm{\lambda}}\left(\bm{\sigma}\right) \right\rangle_{p_{\bm{\lambda}}},
\end{equation}
\noindent where
\begin{equation} \label{eq:pos_phase}
    \left\langle \nabla_{\bm{\lambda}} \mathcal{E}_{\bm{\lambda}}\left(\bm{\sigma}\right) \right\rangle_{\mathbf{D}} = \frac{1}{\vert \mathbf{D} \vert} \sum_{\bm{\sigma} \in \mathbf{D}} \nabla_{\bm{\lambda}} \mathcal{E}_{\bm{\lambda}}\left(\bm{\sigma}\right).
\end{equation}

The expectation value over $p_{\bm{\lambda}}$ poses problems, as one must calculate $Z_{\bm{\lambda}}$ which is computationally costly.
This can be avoided by approximating this expectation value as an estimator over samples from $p_{\bm{\lambda}}$ using Gibbs sampling \cite{hintonPracticalGuideTraining2012}, wherein the variables in a given layer are sampled from its conditional distribution given the current configurations of the other layer: $p_{\bm{\lambda}}\left(\sigma_{id} \vert \bm{h} \right)$ and $p_{\bm{\lambda}}\left(h_j = 1 \vert \bm{\sigma} \right)$.
Starting from an initial configuration of the visible layer (typically an all-zero configuration), $\bm{\sigma}^{(0)}$, one alternates between sampling a new set of \bm{$h$} and $\bm{\sigma}$ via $p_{\bm{\lambda}}\left(\sigma_{id} \vert \bm{h} \right)$ and $p_{\bm{\lambda}}\left(h_j = 1 \vert \bm{\sigma} \right)$.
The number of times one updates the original configuration is referred to as the number of Gibbs sampling steps, $k$.
If one generates a set of samples, $\mathbf{\Gamma}$, in this fashion, the expectation value over the marginal distribution approximates to
\begin{equation} \label{eq:neg_phase}
    \left\langle \nabla_{\bm{\lambda}} \mathcal{E}_{\bm{\lambda}}\left(\bm{\sigma}\right) \right\rangle_{p_{\bm{\lambda}}} \approx \frac{1}{\vert \mathbf{\Gamma} \vert} \sum_{\bm{\sigma}^{(k)} \in \mathbf{\Gamma}} \nabla_{\bm{\lambda}} \mathcal{E}_{\bm{\lambda}}\left(\bm{\sigma}^{(k)}\right).
\end{equation}

To introduce some noise into the gradient updates so as to not get stuck in a local minimum of $C_{\bm{\lambda}}$, stochastic gradient decent, calculating $\left\langle \nabla_{\bm{\lambda}} \mathcal{E}_{\bm{\lambda}}\left(\bm{\sigma}\right) \right\rangle_{\mathbf{D}}$ by way of many averaged updates over mini-batches, $\mathbf{P}$, of $\mathbf{D}$ (equally sized disjoint subsets of $\mathbf{D}$) is performed instead \cite{hintonPracticalGuideTraining2012}.
The individual gradients of $\mathcal{E}_{\bm{\lambda}}$ with respect to the RBM parameters are
\begin{subequations} \label{eq:gradients}
    \begin{align} \label{eq:weight_grad}
        \frac{\partial \mathcal{E}_{\bm{\lambda}}\left(\bm{\sigma}\right)}{\partial W_{ijd}} & = - \mathit{p}_{\bm{\lambda}}\left(h_j = 1 \vert \bm{\sigma} \right)_{j} \sigma_{id} , \\
        \frac{\partial \mathcal{E}_{\bm{\lambda}}\left(\bm{\sigma}\right)}{\partial c_j}     & = - \mathit{p}_{\bm{\lambda}}\left(h_j = 1 \vert \bm{\sigma} \right)_{j},              \\
        \frac{\partial \mathcal{E}_{\bm{\lambda}}\left(\bm{\sigma}\right)}{\partial b_{id}}  & = -\sigma_{id}.
    \end{align}
\end{subequations}
The mini-batch size, $\abs{\mathbf{P}}$, is typically called the \textit{positive-batch size}, while $\abs{\mathbf{\Gamma}}$ is typically called the \textit{negative-batch size}. One full pass through the entire data set $\mathbf{D}$ is called an epoch.
The training hyperparameters used throughout this study for $R = 1.0$ and $1.1$ are given in Tab.~\ref{tab:hyperparameters}.
\begin{table}
    \caption{Hyperparameters used for training.}
    \label{tab:hyperparameters}
    \begin{center}
        \begin{tabular}{ccl}
            \toprule
            Hyperparameter      & Value       & Description         \\
            \hline
            $\eta$              & 0.001, 0.01 & Learning rate       \\
            $\abs{\mathbf{P}}$  & 20          & Positive-batch size \\
            $\abs{\bm{\Gamma}}$ & 10          & Negative-batch size \\
            $k$                 & 10       & \# of Gibbs steps   \\
            \botrule
        \end{tabular}
    \end{center}
\end{table}
A learning rate of 0.001 was employed for the hidden layer scaling, while 0.01 was used for the data size scaling.

\section{Training Threshold} \label{sec:threshold}

Let $\ket{\psi}$ be a normalized state that is an arbitrary superposition of the eigenstates, $\ket{n}$, of a Hamiltonian, $\hat{H}$: $\ket{\psi} = \sum_{n} c_n \ket{n}$. The deviation of the energy of this state, $E_{\psi}$, from the ground state energy of $\hat{H}$, $E_0$, is
\begin{align}
    E_{\psi} - E_0 = & \bra{\psi}\hat{H}\ket{\psi} - \bra{0}\hat{H}\ket{0} = \sum_{n=1}^{\infty} \abs{c_n}^2 \Delta E_n,
\end{align}
where $\Delta E_n = E_n - E_0$.
Relative to the first excited state gap, $\Delta E_1$, the above expression may be written as
\begin{equation}
    \frac{E_{\psi} - E_0}{\Delta E_1} = \abs{c_1}^2 + \sum_{n=2}^{\infty} \abs{c_n}^2 \frac{\Delta E_n}{\Delta E_1}.
    \label{eq:energy_difference}
\end{equation}
When $E_{\psi}$ approaches $E_0$, we expect the dominant contribution to the difference to be from the first excited state.
Therefore, neglecting all terms beyond the first on the right-hand side of Eq.~\eqref{eq:energy_difference}, we see that the left-hand side should act as a proxy for the amount of excited state contamination in $\ket{\psi}$.

\bibliographystyle{apsrev4-1}
\bibliography{refs}

\end{document}